# The Wide Field Imager Instrument for Athena

Norbert Meidinger*[a], Josef Eder[a], Tanja Eraerds[a], Kirpal Nandra[a], Daniel Pietschner[a], Markus Plattner[a], Arne Rau[a], and Rafael Strecker[a]

[a]Max-Planck-Institut für extraterrestrische Physik, Giessenbachstrasse, 85748 Garching, Germany

**ABSTRACT**

The WFI (Wide Field Imager) instrument is planned to be one of two complementary focal plane cameras on ESA's next X-ray observatory Athena. It combines unprecedented survey power through its large field of view of 40 amin x 40 amin together with excellent count rate capability ($\geq$ 1 Crab). The energy resolution of the silicon sensor is state-of-the-art in the energy band of interest from 0.2 keV to 15 keV, e.g. the full width at half maximum of a line at 7 keV will be $\leq$ 170 eV until the end of the nominal mission phase. This performance is accomplished by using DEPFET active pixel sensors with a pixel size of 130 µm x 130 µm well suited to the on-axis angular resolution of 5 arcsec half energy width (HEW) of the mirror system. Each DEPFET pixel is a combined sensor-amplifier structure with a MOSFET integrated onto a fully depleted 450 µm thick silicon bulk. Two detectors are planned for the WFI instrument: A large-area detector comprising four sensors with a total of 1024 x 1024 pixels and a fast detector optimized for high count rate observations. This high count rate capable detector permits for bright point sources with an intensity of 1 Crab a throughput of more than 80% and a pile-up of less than 1%. The fast readout of the DEPFET pixel matrices is facilitated by an ASIC development, called VERITAS-2. Together with the Switcher-A, a control ASIC that allows for operation of the DEPFET in rolling shutter mode, these elements form the key components of the WFI detectors. The detectors are surrounded by a graded-Z shield, which has in particular the purpose to avoid fluorescence lines that would contribute to the instrument background. Together with ultra-thin coating of the sensor and particle identification by the detector itself, the particle induced background shall be minimized in order to achieve the scientific requirement of a total instrumental background value smaller than 5 x $10^{-3}$ cts/cm$^2$/s/keV. Each detector has its dedicated detector electronics (DE) for supply and data acquisition. Due to the high frame rate in combination with the large pixel array, signal correction and event filtering have to be done on-board and in real-time as the raw data rate would by far exceed the feasible telemetry rate. The data streams are merged and compressed in the Instrument Control and Power distribution Unit (ICPU). The ICPU is the data, control and power interface of the WFI to the Athena spacecraft. The WFI instrument comprises in addition a filter wheel (FW) in front of the camera as well as an optical stray-light baffle. In the current phase A of the Athena project, the technology development is performed. At its end, breadboard models will be developed and tested to demonstrate a technical readiness level (TRL) of at least 5 for the various WFI subsystems before mission adoption in 2020.

## 1. INTRODUCTION

The Advanced Telescope for High ENergy Astrophysics (Athena)[1],[2] of the European Space Agency (ESA) is planned to be equipped with a single large-aperture X-ray mirror assembly[3] which images X-ray photons onto one of two complementary and interchangeable focal plane instruments:
The X-ray Integral Field Unit (X-IFU) provides very high spectral resolution by using transition edge sensors operated at cryogenic temperatures[4]. The Wide Field Imager (WFI) provides a detector covering a large field of view of 40 amin x 40 amin and a detector featuring high count rate capability for the observation of very bright point sources.

---

* nom@mpe.mpg.de,  phone: ++49 89 300003866,  fax: ++49 89 300003569,  mpe.mpg.de

For both purposes, silicon active pixel sensors of DEPFET (depleted p-channel field effect transistors) type are used. Such a detector shows good spectral resolution over the required broad energy band from 0.2 keV to 15 keV, e.g. meet the requirement of a FWHM(7 keV) ≤ 170eV throughout the mission, and facilitates high time resolution, especially 80 μs for the high count rate capable detector. Both, the large field of view and the high count rate capable detector, are based on the same concept but will be tailored specifically to their individual tasks. The key science drivers for the Athena WFI instrument are described in detail in ref.[5].

An overview about the WFI instrument with its subsystems is presented in the next section. The camera head (CH) with focal plane detectors is described in section 3. The concept for the signal chain and electronics of WFI are explained in section 4. A low instrumental background is of high importance for the observation of faint extended sources. The analysis how to achieve the required low particle induced background is presented in section 5. The schedule for the WFI instrument development concludes the description of the WFI camera for the Athena project.

## 2. WFI OVERVIEW

### 2.1. Conceptual design

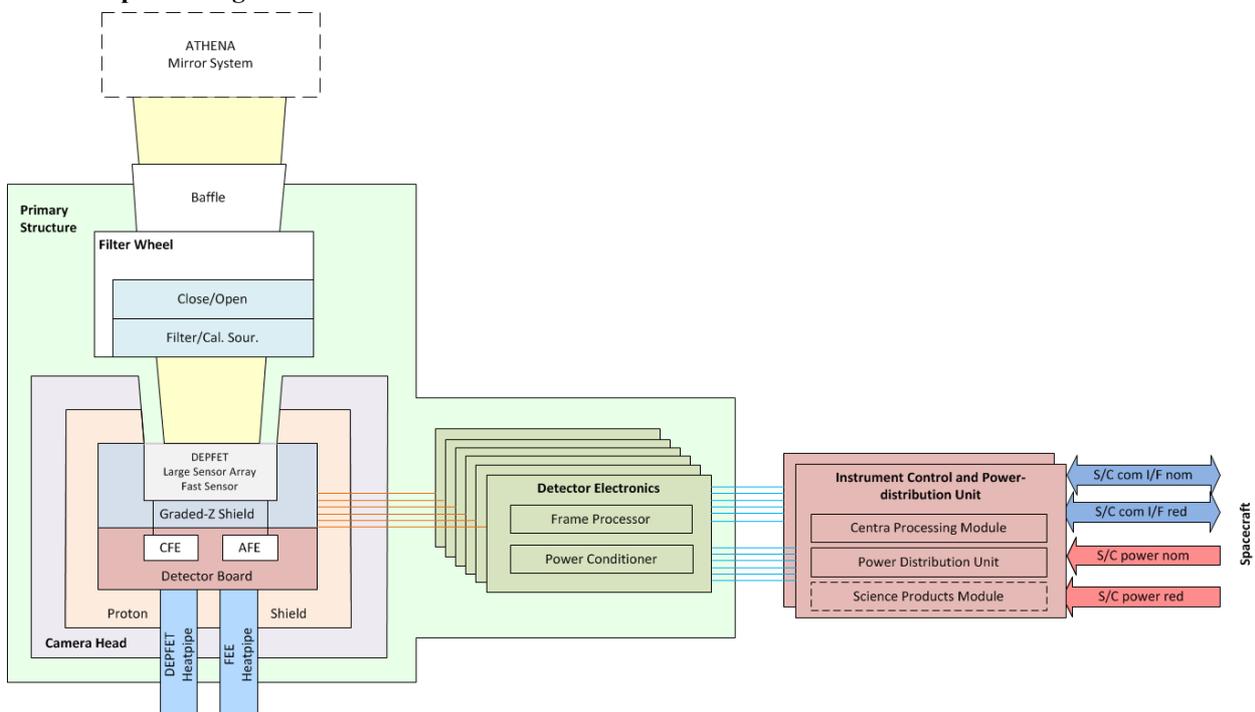

Figure 1: The block diagram of the WFI instrument shows the subsystems: filter wheel with optical stray-light baffle in front of it, camera head comprising the detectors and the six detector electronics. All these subsystems are mounted on a primary structure, which is accommodated on the focal plane module (FPM) of the satellite. Accommodated on the FPM are also the two WFI instrument control and power distribution units (ICPUs) and the radiator panels. The latter ones cool the subsystems via heat pipe links to the temperature required for the components of the subsystem. The detectors with the sensors, control and analogue frontend electronics (AFE, CFE) as main components are surrounded by a graded shield whose outer layer is the proton shield.

Figure 1 gives an overview about the various subsystems of the WFI instrument. A filter wheel offers four slots for the required various functionalities:

i) An UV and visible light blocking filter is needed as the X-ray sensor is also sensitive to UV and visible light, which is emitted by the observed source and transmitted through the Athena optics. The by UV and visible light generated electrons in the silicon sensor would merge with the X-ray signal electrons and degrade the energy resolution. A suppression of visual photons by about 7 orders of magnitude is envisaged. The blocking filter will be split in two parts: one part will be directly deposited on the photon entrance window of the sensor (90 nm Al) and the other part is a foil (e.g. 40 nm Al on 200 nm polyimide substrate) mounted in the filter wheel. For the large field of view detector a filter of 160 mm x 160 mm size is needed. The high count rate capable detector requires a size of 15 mm x 15 mm due to the distance of the filter wheel disk to the detector and the divergence of the imaged X-ray photons. The filter foil shall be supported by a mesh, as the WFI instrument is not planned to be accommodated inside a vacuum chamber. The filters will thus be exposed to the acoustic loads of the satellite launch.

ii) An open position is needed for efficient evacuation of the filter wheel and camera head during testing in vacuum chambers and finally in space. This position permits also special observations with high quantum efficiency at low X-ray energies if optical loading is sufficiently small.

iii) A closed position is available for sensor protection and instrumental background measurements.

iv) The fourth position hosts the on-board calibration sources. They facilitate recalibration of the camera during the mission, in particular during observations with the X-IFU instrument.

Further details about the filters and the filter wheel are given in ref.[6] and [7]. The camera head is mounted directly to the filter wheel (see Figure 2). Both are under ambient pressure in the baseline design, i.e. no vacuum chamber encloses the sensitive components sensors and filters. This decision is driven by the high complexity of a vacuum chamber solution in consequence of the large number of heat pipes and electrical interfaces.

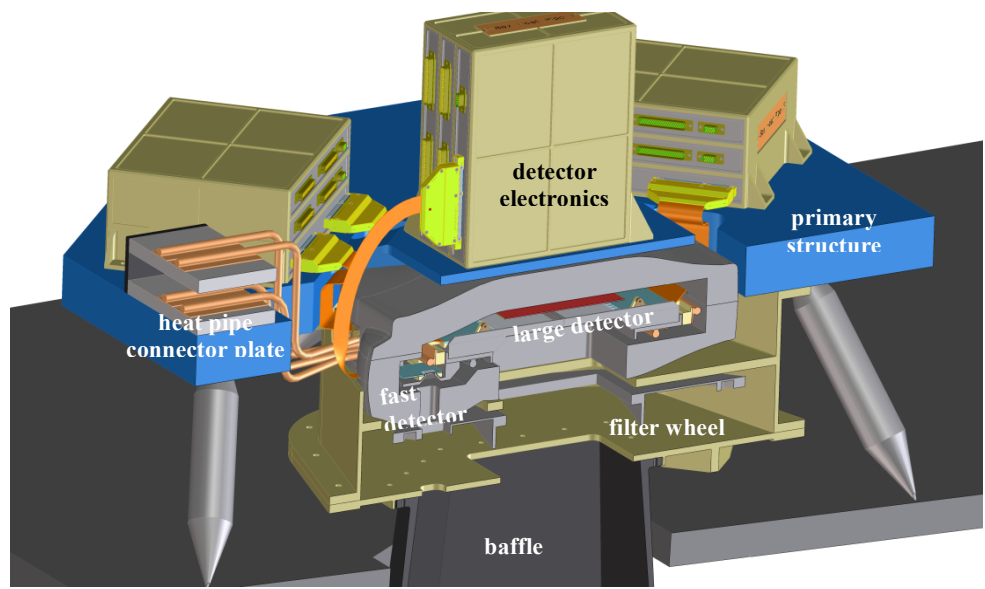

Figure 2: Section drawing of the WFI instrument. The camera head in the centre consists of the large field of view detector and the defocused fast high count rate capable detector to the left of it. The detectors are surrounded by the graded-Z shield, which suppresses the fluorescence radiation in order to minimize the instrumental background. The outermost layer is the proton shield, which reduces the proton flux to the detector in order to mitigate the radiation damage. The filter wheel is mounted in front of the camera head together with the baffle towards the Athena mirror system. The primary structure carries this assembly and furthermore the six detector electronics subsystems. The connector plate mounted but thermally insulated from the primary structure is shown on the left hand side. It is the thermal interface between the camera head internal heat pipes and the external heat pipes that connect WFI to the radiator panels on the focal plane module (FPM), which are not shown here. The primary structure is mounted with four bipods to the FPM on which the other focal plane camera, the X-IFU, is mounted too. The position of the ICPU boxes (nominal and redundant one) underneath the FPM can be chosen arbitrarily.

Six detector electronics are needed for the four quadrants of the large field of view detector and the two separately operated halves of the fast detector. Each detector unit is connected to the DE by flexible leads. The DE supplies the detector with voltage and timing signals and digitizes the analogue detector output signals. The pre-processed six data streams from the detector electronics are merged, analysed and further compressed in the instrument control and power distribution unit (ICPU). The ICPU accommodates potentially an on-board data analysis unit called science products module and the controller for the filter wheel and is responsible for the thermal control system of WFI. In order to increase the instrument reliability, a nominal and redundant ICPU are envisaged in the WFI concept. Further details on the WFI subsystems can be found in ref.[8] and [9].

## 3. CAMERA HEAD WITH FOCAL PLANE DETECTORS

The WFI uses active pixel sensors (APS) of DEPFET type[9]. They measure precisely the energy, incidence position and arrival time of the X-ray photons focused by the Athena mirror system to the detector with high quantum efficiency. DEPFET APS are actually a combined sensor and amplifier structure. It is a back-illuminated sensor like the PNCCDs used for XMM-Newton and eROSITA with 450 µm depletion layer. However, in contrast to the CCD, each pixel is equipped with a transistor for sensing the stored signal charge and a second transistor for clearing the charge after readout. The signal electrons generated in the silicon sensor by an incoming X-ray photon are collected in the internal gate below the sensing transistor and increase the current proportional to the number of charges.
Although DEPFETs were already developed for the MIXS[10] instrument on-board of ESA's BepiColombo satellite to Mercury, an optimized DEPFET transistor geometry and an enhanced fabrication technology is needed for the WFI instrument because of the challenging requirements of the project, in particular with respect to sensor area and time resolution. Prototype devices have been designed and are currently produced in the semiconductor laboratory of Max-Planck-Society. Testing a large variety of these DEPFET prototypes for WFI with pixel arrays from 64x64 up to 256x256, 512x64 and 128x512, will provide the necessary information for the decision of transistor design and technology option for the Athena flight DEPFET sensors.

For DEPFET readout, an analogue signal processor called Veritas-2 is under development[11]. The fully differential ASIC using trapezoidal shaping provides 64 parallel channels. In case of a large quadrant DEPFET, eight Veritas-2 ASICs are needed for readout of a DEPFET line with 512 pixels. During signal processing of the pixels, the stored signals of the previous line are multiplexed to the readout node of the ASIC in order to minimize readout time. A quadrant needs thus only eight ADCs in the detector electronics. A special feature of Veritas-2 is that it offers two readout modes, the source follower mode and the drain current readout mode. Both have been tested with promising results[12]. While the drain current readout method facilitates faster readout, the source follower mode is more tolerant to transistor current fluctuations over the sensor area. Both methods will be further studied with the DEPFET prototype devices for WFI.

The DEPFET is operated in rolling shutter mode: one DEPFET line is switched on while the others accumulate photons without power consumption. This is the trade-off between high readout rate as all pixels of a row are read out simultaneously and a moderate power dissipation of the detector. For this purpose of controlling the DEPFET, an ASIC has been developed called Switcher. According to the input programme sequence, the external gates of the DEPFETs of the active line are switched to a more negative voltage to switch the transistor on for readout of the signal currents which depends on the number of electrons stored in the internal gate of the DEPFETs. After readout, the charge in the internal gate is cleared during a short period of <1 µs by switching the gate of the clear transistor and its drain to a more positive voltage level. At the end of the readout of the line, the voltage applied to the external gate is switched back to the original voltage that switches the transistors off again. Each Switcher-A ASIC developed for WFI can control a block of 64 lines and thus eight Switcher-A ASICs are needed for complete control of a WFI quadrant with 512 lines.

### 3.1. Focal plane layout

The WFI camera comprises two focal plane detectors. One provides a large field of view of 40 amin x 40 amin and the other one features a high count rate capability for bright point sources (see Figure 3). Also they are quite different in size, both have in common a pixel size of 130 µm x 130 µm (corresponding to an angular element of 2.2 arcsec x 2.2 arcsec) which allows a sufficiently accurate source position reconstruction for a point spread function (PSF) of 5 arcsec HEW (or 3 arcsec as goal) of the mirror system[8]. All sensors, four in case of the large field of view detector and two sensor halves of the fast detector, are read out in rolling shutter mode. The Athena mirror system will point either to one of the two WFI detectors, the large field of view or the high count rate capable one, or to the X-IFU detector depending on the source and the associated observation request.

### 3.1.1. Large-area sensor assembly

The large field of view detector has to span a 40 amin x 40 amin large field of view for which 1024 x 1024 pixels are necessary with the given pixel size. As the physical size just of the total pixel area is already more than 13 cm in square but the wafers diameters available for DEPFET sensor fabrication have a diameter of only 15 cm, the field of view is subdivided into four quadrants (see Figure 3). Therefore, insensitive regions appear between the sensitive pixel areas of the four DEPFET sensor chips. Furthermore, a structure called DEPFET frame is needed surrounding each DEPFET sensor to give the necessary mechanical support and thermal coupling to the heat pipes. The resulting insensitive area can be compensated by observing with an appropriate dithering pattern.

Each of the four identical quadrant sensors with 512 lines and 512 columns will be controlled by 8 Switcher-A ASICs and read out by eight Veritas-2 ASICs. The quadrants are independent of each other. The time resolution for the large detector is required to be <5 ms. The four quadrants are integrated into a structure frame, which connects them mechanically and thermally (see Figure 3).

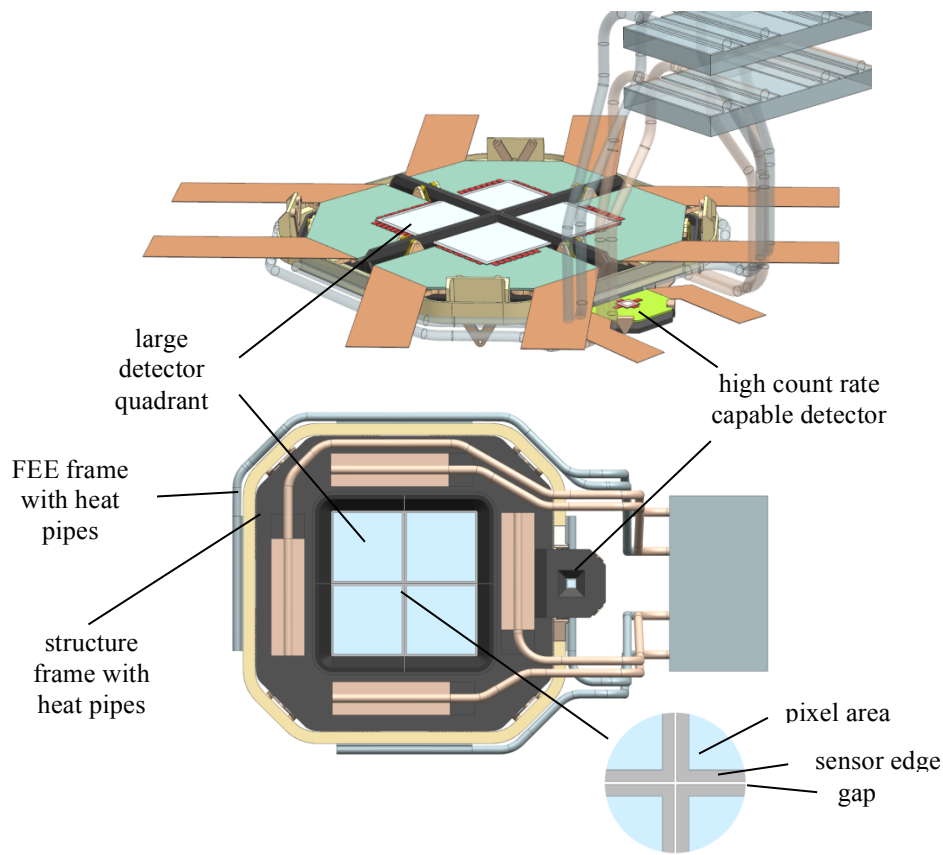

Figure 3: Focal plane design of the WFI instrument with view on the structured side with the ASICs (top picture) and on the photon entrance window side (bottom picture). The frontend electronics boards with the flexible leads are shown as well as the heat pipes needed for cooling the sensor and the frontend electronics respectively. The insensitive area between the quadrants will be mainly caused by the edges of the sensors, which are mounted to support frames, and not by the gap between the quadrants.

### 3.1.2. High count rate capable sensor

The sensor for the high count rate capable WFI detector is a DEPFET matrix with 64 lines and 64 channels[8]. It is split into two identical halves for an operation in split full frame mode. This means control and readout of the two halves in parallel, which gains a factor of two in time resolution. The requirements for the time resolution of the high count rate capable detector is 80 µs. The required throughput of >80% and pile-up <1% for a point source with 1 Crab intensity can be achieved by defocusing the detector by about 35 mm[8]. For this purpose, the detector is mounted defocused in contrast to the large field of view detector (see Figure 3).

In the course of the detector development, various concepts have been studied for the high count rate capable detector. The basic groups are non-gateable DEPFETs as planned for the large detector and gateable DEPFETs[13]. Although the concepts for gateable DEPFETs with additional signal storage region per pixel are quite promising, their advantage for the spectral response of the high count rate capable detector by suppressing energy misfits occurring during the readout is not crucial for the Athena science and can be modeled well as part of the detector response matrix. From the operational point of view, these devices show a relatively small and thus critical window for the operating voltages[14].

Consequently, a similar pixel design as envisaged for the large detector will be applied for the high count rate capable detector causing synergy effects for the development of the two WFI detectors.

### 3.2. Frontend electronics

Each sensor assembly (i.e. of the four quadrants and the high count rate capable sensor) requires dedicated frontend electronics (FEE). Both form an inseparable unit as they are connected by a large number of wire bonds between the DEPFET sensor on one side and the frontend printed circuit board (PCB) with the ASICs on the other side. The wire bonds between the two parts require a rigid mechanical structure of such a detector subunit. Any movements of the bond wires caused by cooling of the sensor or vibrations due to the satellite launch have to be minimized. However, the thermal coupling shall be as low as possible on the other hand as the sensor requires a relatively low temperature between -80°C to -60°C while the ASICs shall operate at a considerably higher temperature. Without thermal decoupling, a larger radiator area would be needed on the spacecraft.

A prototype version of the Veritas-2 ASIC is presently under test for readout of the DEPFET sensor. Both readout options, source follower readout and drain readout, showed promising results[12]. An already existing version of the control Switcher ASIC has been redesigned to meet the requirements of the WFI instrument, in particular two-side buttability of the quadrants. As soon as the production of the DEPFET sensors is finished, experimental tests will be performed. Both ASICs are produced with CMOS technology, in case of the Veritas-2 it is AMS 0.35µm 3.3V CMOS technology and for the Switcher-A an AMS 0.35µm 50V CMOS process is used. For each of the four quadrants, a multi-layer rigid-flex frontend electronics PCB is projected comprising about 400 lines in the flexible leads.

### 3.3. Detector assembly

Rigid mechanical support structures are needed for the assembly of the detector subunits to the two WFI detectors. Each subunit consists of a sensor assembly and a frontend electronics. Furthermore, optimum thermal coupling between the sensor assemblies is necessary, which is provided by the detector structure frame. The FEE frame accomplishes the same for the frontend electronics assemblies that are operated at a higher temperature (see Figure 3). The mechanical connection but thermal insulation between structure frame and FEE frame is accomplished by suitably shaped PEEK links. Heat pipes are mounted at the top of each frame, which thermally couple it via the connector plate on the WFI primary structure to the respective radiator panels facing the cold space.
The electrical interfaces within the detector assembly are accomplished by aluminum wire bonds:
   a) the DEPFET transistors (source or drain contact dependent on the type of finally selected readout mode) are connected column-wise to the Veritas-2 input stages.
   b) the DEPFET external gate, clear gate and clear contact are connected line-wise to the three corresponding Switcher-A ports.
   c) the supply voltage contacts on top and bottom side of the DEPFET sensor chip are connected to the corresponding supply lines on the PCB.
The other electrical interface between frontend PCB and detector electronics is accomplished by flexible leads (the flexible part of the FEE PCB) with connectors.

First tests with small detector breadboards and laboratory electronics revealed an energy resolution FWHM(5.9 keV) of ≤140 eV for a moderate readout time of 8 µs and ≤150 eV for a faster readout of 4 µs per line[12]. When the presently produced Athena prototype DEPFETs will be available for tests, improvements in energy resolution and readout time are expected. The scientific requirements for energy resolution are a FWHM of ≤80 eV at 1 keV and a FWHM of ≤170 eV at 7 keV until end of nominal life of the Athena mission.

### 3.4. Thermo-mechanical design

The detector electronics has to be accommodated in direct vicinity of the detectors in order to keep the length of the flexible lead at a minimum (see Figure 2). This is important as the detector signals are already amplified but still analogous and a degradation has to be avoided. The six detector subunits are thus surrounded by six detector electronics. The detectors are enclosed in the camera head providing the necessary mechanical support, shielding and thermal environment. Cut-outs in the camera head are necessary for the feedthroughs of heat pipes and flexible leads, which motivated the approach to develop a design without vacuum chamber. The filter wheel with optical stray light baffle is mounted in front of the camera head. All these components are integrated on the primary structure which has four bipods as interface to the focal plane module of the Athena spacecraft.

Radiation damage to the detector, in particular to the sensors has to be minimized by shielding of protons and alpha particles. While the detector side opposite to the mirror system can be shielded relatively easily, the shielding of the photon entrance side requires a more complex geometry because of the X-ray photon aperture. This has been considered in the design of the filter-wheel and the optical stray-light baffle shown in Figure 2. Apart from shielding, a sufficiently low operating temperature of the sensor mitigates the dark current increase which is the dominant radiation damage effect. A temperature region between -80°C and -60°C is envisaged for operation of the silicon sensor in order to reduce the thermal generation current arising from radiation damage and also potential micrometeoroid impacts. The front-end electronics shall be operated at a higher temperature range (T=-20°C to 0°C) but is thermally coupled to the sensors by more than 8700 bond wires. Hence, the thermal gradient between sensor and front-end board has to be limited. Passive cooling is sufficient for both units and heat pipes will establish the thermal link to the dedicated radiator panels on the FPM of the satellite. The same is valid for the detector electronics, which is actually the most power consuming and heat dissipating subsystem but can be operated at relatively warm temperature. Details about the thermal concept are presented in ref.[15].

### 3.5. WFI camera requirements and characteristics

Table 1 gives an overview of the main requirements and performance parameters of the WFI instrument. Further information in particular about the filters and the quantum efficiency can be found in ref.[6] and [8].

Table 1: WFI detector characteristics

| Parameter | Characteristics |
|---|---|
| Energy range | 0.2 keV – 15 keV |
| Pixel size | 130 µm x 130 µm (corresp. to 2.2" x 2.2") |
| Operating mode | rolling shutter |
| Operating time | nonstop operation possible |
| Large field of view detector | 40' x 40' <br> 1024 x 1024 pixel (4 quadrants) |
| High count rate capable detector | 1 Crab: >80% throughput and <1% pile-up <br> 64 x 64 pixel (split full frame mode) <br> FoV =143" x 143"; mounted defocused |
| Energy Resolution | FWHM(1 keV) $\leq$ 80 eV (end of life) <br> FWHM(7 keV) $\leq$ 170 eV (end of life) |
| Time Resolution (full frame) <br> Fast detector <br> Large detector | <br> 80 µs <br> <5 ms |
| Quantum efficiency <br> (on-chip + ext. filter) | 20% @ 277 eV <br> 80% @ 1 keV <br> 90% @ 10 keV <br> transmission of visible light $\approx 10^{-7}$ |
| Non X-ray background (L2 orbit) | $< 5 \times 10^{-3}$ cts cm$^{-2}$ s$^{-1}$ keV$^{-1}$ <br> in 2 keV-7 keV band for 60% of observing time |

## 4. SIGNAL CHAIN AND ELECTRONICS

The output of Veritas-2 provides fully differential detector signals. These analog signals are routed from CH to DE via the flexible leads. Inside DE, a 14-bit ADC digitizes the analogue data. For this reason, the length of the flexible lead has to be kept short and the detector electronics has thus to be accommodated close to the detector. After digitization of the signals, corrections of the offset and common mode of the detector signals are performed by the frame processor, presumably based on a Virtex-5 FPGA. In the next step, the events are selected by applying a lower and upper threshold. This rejects non-photon events caused by electronic noise and particles. The extracted pixel signals can then be checked with respect to validity of the event pattern. Only events that are spread over up to 2x2 pixels are valid according to the charge collection process in the silicon sensor chip and can be accepted for spectroscopy. If the event area exceeds 2x2 pixels, pattern pile-up has occurred and the energy of the individual events cannot be determined. This DE internal pre-processing reduces the data rate before the event list is finally sent to the ICPU. In the ICPU, the data streams of the six detector subunits are merged. After further analysis, the data are compressed and sent to the mass memory on the satellite. During the contact times of the satellite with the ground stations, the data are dumped down and the near real-time analysis as well as the standard analysis can be performed.

Most challenging here is the real-time pre-processing of the detector data in the detector electronics. The rate is about 209 Mpixels/s for the large field of view detector (assuming a 5 ms frame period) and 51Mpixel/s for the high count rate capable detector. This pixel pre-processing rate is more than a factor of 100 higher compared to that of the standard mode of the EPIC PN camera on-board of XMM-Newton. Further details can be found in ref.[8] and [16].

## 5. INSTRUMENTAL BACKGROUND

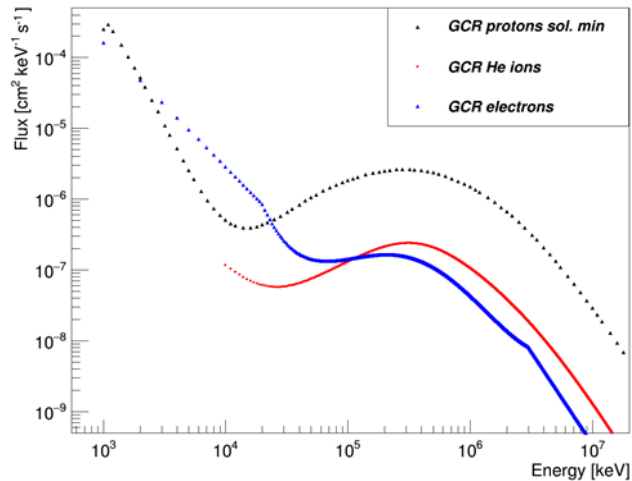

Figure 4: Input spectra for the background simulation for cosmic-ray protons, electrons and Helium ions. The proton spectrum corresponds to CREME 96 standard model for solar minimum[20], the electron and He ion spectrum are taken from ref.[21].

Many of the observations performed with the Athena WFI instrument require a low background level. Two main types of particle induced instrumental background components can be distinguished: electrons and photons generated by high-energetic cosmic particles impinging on material surrounding the WFI DEPFETs and background due to soft protons, which enter through the mirror optics. The first contribution includes the generation of fluorescence lines in the vicinity of the detectors. For their suppression, both WFI detectors shall be surrounded by a graded Z-shield with decreasing atomic number Z from outer to inner layers. By appropriate design, the X-ray photons generated in the camera will be absorbed in the next inner layer until finally mostly very low-energy Auger electrons are emitted in the innermost layer with very low Z. Such an approach has already been made for the eROSITA cameras[17].

According to the Athena scientific requirement, the instrumental background should be below $5 \times 10^{-3}$ cts/cm$^2$/s/keV. Monte Carlo simulations, employing the GEANT4-based[18] GRAS tool[19], are used to estimate the expected instrumental background at Lagrange point L2. Main contributions to the instrumental background are generated by cosmic ray protons, electron and Helium ions. The input spectra used for the simulations are presented in Figure 4. The cosmic ray proton spectrum follows the CREME96[20] convention for solar minimum; the Helium ion and electron spectra follow the formula given in ref. [21] for mission averaged cosmic particles.

The simulations use a simplified mass model of the WFI camera head, filter wheel and baffle design. It will be updated when major modifications occur in the course of the instrument development process. The applied simplifications help to keep the computing time under control and allow for an easier control of the dependence of the background spectra on the mass model. By comparing slightly modified versions of the mass model, it has been found out that the inclusion of coatings on the DEPFET sensor into the simulation mass model has a significant effect on the background spectra. The considered coatings are a 3 µm BCB (Benzocyclobutene) thick layer on the off-mirror side of the detector, which serves for passivation of the sensor chip and a 90 nm thick Aluminum coating on the mirror side of the detector, which serves as on-chip filter to reduce the contamination of the x-ray signal by optical light. The coatings reduce the background contribution of low-energy secondary electrons, as shown in the comparison of the two plots in Figure 5. These plots show the instrumental background, i.e. energy stored in adjacent hit pixels due to cosmic ray protons, once in the case no coating is considered and once in the case the coatings have been taken into account. Low-energy electrons generated in the material surrounding the detectors are absorbed in the coating layers. Secondary electrons generated in the coating however are likely to hit the detector close to the particle, which produced them. Hits are thus created in the detector, which can be discarded from the background spectra due to their direct vicinity to pixels with energy deposition above the signal energy range.

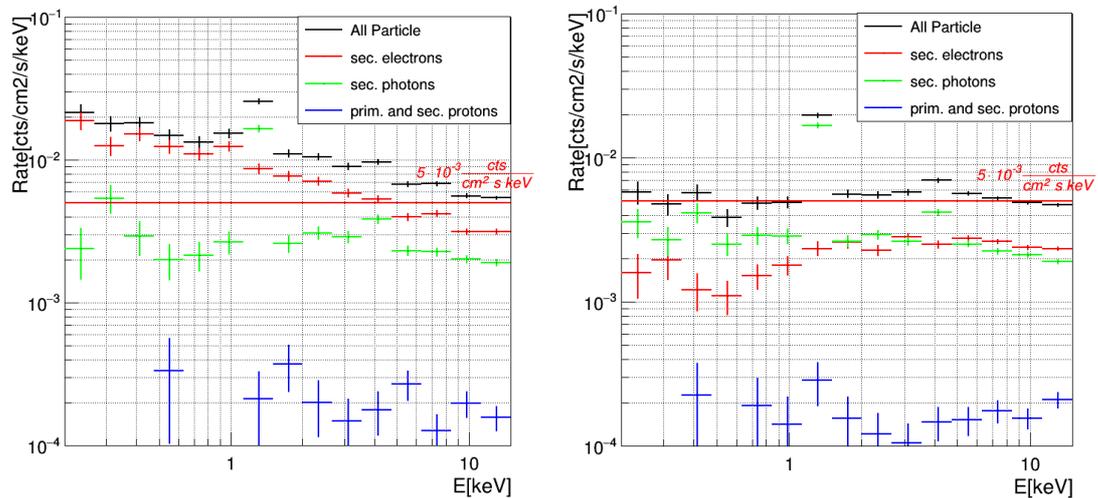

Figure 5: Instrumental background due to cosmic-ray protons, on the left hand side in case no detector coating is included in the mass model of the simulation and on the right hand side in case 3 µm BCB coating on the off-mirror side and 90 nm Al coating on the photon entrance window side is included in the mass model of the simulation. The total background due to cosmic ray protons is marked in black, the contribution due to secondary electrons hitting the detector is marked in red, that due to secondary photons in green and that due to primary and secondary protons in blue; no further event selection is applied apart from recombination of pixel hits to events and rejection of pixels with energies above 15keV and their direct vicinity.

In order to further reduce the remaining background, several analysis strategies have been tested. One of them is the rejection of readout frames in which more energy than possible by a 15 keV X-ray photon has been deposited in at least one pixel. The idea behind this analysis strategy is that secondary particles produced by high energy particles are likely to hit the detector close to the particle which produced them, thus leading to signal-like-hits generated by low energetic secondary electrons and photons close to pixels with energy deposition >15 keV. The method is especially effective if the secondaries are generated on a surface parallel and close to the detector. Table 2 shows background reduction and signal loss once for a minimum readout time of 1.3 ms and once for the readout time of 5 ms according to the scientific requirements. Increasing the readout time reduces both, signal and background hits, as both signal and background events have a higher likelihood to be detected during a rejected frame. The special implications of the rolling shutter readout mode have been taken into account for this analysis.

The final background spectrum after the application of the described selection method is presented in Figure 6 for a readout time of 1.3 ms. Please note that the graded Z-shield was incomplete in the simulation model and thus the Al-K fluorescence lines still appears at 1.5 keV energy. Taking this into account, the background level is about in agreement with the scientific requirement of <5 x $10^{-3}$ cts/cm$^2$/s/keV.

Table 2: Background reduction, signal (event) loss, change in signal over background and signal over $\sqrt{\mathbf{background}}$ given once for a time resolution of 1.3 ms and 5 ms respectively. The here applied selection method is the rejection of a quadrant of a readout frame if in this quadrant for at least one pixel an energy deposition higher than 15 keV is observed during this readout period.

| Time resolution | Background reduction | Signal loss | Change in Signal/Background | Change in $\frac{S}{\sqrt{B}}$ |
|---|---|---|---|---|
| 1.3 ms | 39% | 18% | 1.3 | 1.0 |
| 5 ms | 73% | 64% | 1.3 | 0.7 |

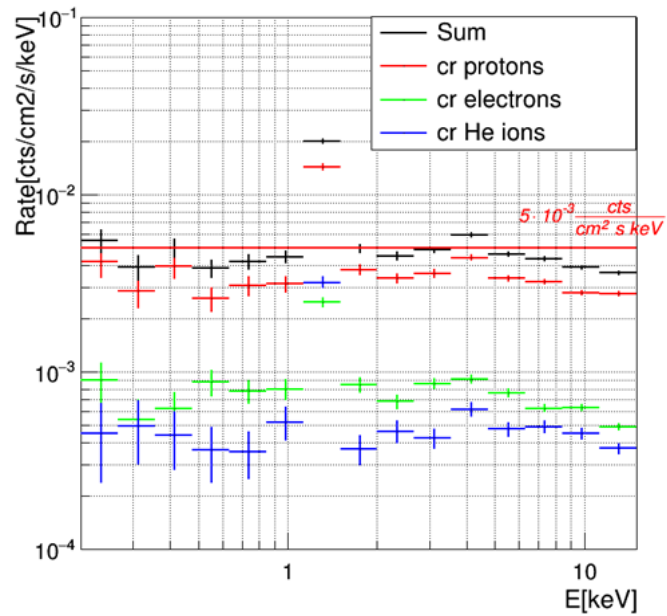

Figure 6: Instrumental background of WFI due to cosmic-ray protons (red), cosmic-ray electrons (green) and cosmic-ray Helium ions (blue). The WFI mass model has been taken into account with sensor coating and the rejection of quadrant readout frames if in this quadrant in at least one pixel an energy deposition higher than 15 keV is observed. A readout time of 1.3 ms has been chosen for the simulation. Please note, the Al-K fluorescence line at 1.5 keV appears here still as in the applied mass model no complete graded-Z shield has been considered which will suppress it.

## 6. WFI PROJECT SCHEDULE

The WFI instrument follows a classical model philosophy for a space project. The instrument concept for the Athena mission is presently in the so-called technology development activity phase. At the end of this phase, breadboard models shall demonstrate a technical readiness level (TRL) of at least 5 for the WFI subsystems. Critical elements are the DEPFET detector, the high-speed signal pre-processing in the detector electronics in real-time and the ultra-thin, large-area optical blocking filter because of the susceptibility to damage due to acoustic noise. After this phase, a preliminary design of the WFI instrument will be developed and reviewed in 2020. Subsequently, an engineering model, a structural and thermal model and an electrical functions model will be developed and tested. The qualification model follows the critical design review. Finally, the flight model will be assembled, tested, calibrated and integrated into the FPM. The launch of the Athena satellite is planned for end of 2028.

## 7. SUMMARY AND OUTLOOK

The WFI instrument for Athena will provide two complementary detectors: a detector with a large field of view of 40 amin x 40 amin and a spatial resolution well suited to the on-axis angular resolution of 5 arcsec HEW of the mirror system. The other detector permits high count rate capability with <1% pile-up at 1 Crab source intensity and high time resolution of 80 μs. Both detectors utilize active pixel sensors of DEPFET type and have the requirement to show excellent state-of-the-art energy resolution, e.g. FWHM(1 keV) < 80 eV, until the end of the mission. The WFI is furthermore devised to show a very low particle-induced instrumental background of <5 x $10^{-3}$ cts/cm$^2$/s/keV. Although the development of the WFI instrument is a challenge, in particular with respect to sensor area, time resolution and assembly, integration & test, a straightforward conceptual design has been created to facilitate the progress of the project. Finally, the WFI satellite instrument will be the result of the development work accomplished by the international WFI consortium.

## ACKNOWLEDGMENTS

The authors are grateful to all colleagues and institutions that supported the Wide Field Imager instrument for Athena. The work was funded by the Max-Planck-Society and the German space agency DLR (FKZ: 50 QR 1501). Development and production of the DEPFET sensors for the Athena WFI is performed in collaboration between MPE and the MPG Semiconductor Laboratory (HLL).